\batchmode
\makeatletter
\def\input@path{{\string"E:/Trabajo Angel/Mis articulos/Finished/Fusion 2019/MBM filter/\string"}}
\makeatother
\documentclass[british,english]{IEEEtran}
\usepackage[T1]{fontenc}
\usepackage[latin9]{inputenc}
\usepackage{float}
\usepackage{amsmath}
\usepackage{amssymb}
\usepackage{graphicx}

\makeatletter

\floatstyle{ruled}
\newfloat{algorithm}{tbp}{loa}
\providecommand{\algorithmname}{Algorithm}
\floatname{algorithm}{\protect\algorithmname}

\usepackage{cite} 
\usepackage[margin=8pt,font=footnotesize]{caption}
\usepackage{algorithm}
\usepackage{algpseudocode}
\usepackage{amsmath}  % You need this for the math
\allowdisplaybreaks

\@ifundefined{showcaptionsetup}{}{%
 \PassOptionsToPackage{caption=false}{subfig}}
\usepackage{subfig}
\makeatother

\usepackage{babel}
\addto\captionsbritish{\renewcommand{\algorithmname}{Algorithm}}

\begin{document}

\title{Gaussian implementation of the multi-Bernoulli mixture filter}

\author{Ángel F. García-Fernández$^{\star}$, Yuxuan Xia$^{\circ}$, Karl
Granström$^{\circ}$, Lennart Svensson$^{\circ}$, Jason L. Williams$^{\dagger}$\\
{\normalsize{}$^{\star}$Dept. of Electrical Engineering and Electronics,
University of Liverpool, United Kingdom}\\
{\normalsize{}$^{\circ}$Dept. of Electrical Engineering, Chalmers
University of Technology, Sweden}\\
$^{\dagger}${\normalsize{}Commonwealth Scientific and Industrial
Research Organisation, Australia}\\
{\normalsize{}Emails: angel.garcia-fernandez@liverpool.ac.uk, firstname.lastname@chalmers.se,
jason.williams@data61.csiro.au}}

\maketitle
\thispagestyle{empty}
\begin{abstract}
This paper presents the Gaussian implementation of the multi-Bernoulli
mixture (MBM) filter. The MBM filter provides the filtering (multi-target)
density for the standard dynamic and radar measurement models when
the birth model is multi-Bernoulli or multi-Bernoulli mixture. Under
linear/Gaussian models, the single target densities of the MBM mixture
admit Gaussian closed-form expressions. Murty's algorithm is used
to select the global hypotheses with highest weights. The MBM filter
is compared with other algorithms in the literature via numerical
simulations.
\end{abstract}


\section{Introduction}

Multiple target tracking (MTT) is an important problem in many applications,
such as, surveillance, autonomous vehicles and air traffic control
\cite{Blackman_book99,Granstrom18b}. Relevant MTT algorithms are
multiple hypothesis tracking \cite{Reid79,Kurien_inbook90,Coraluppi14,Brekke18},
joint probabilistic data association \cite{Fortmann83}, and algorithms
based on random finite sets (RFSs) \cite{Mahler_book14}. 

In the RFS formulation, the (multi-target) filtering density contains
the information of the target states at the current time step. This
density can be used to estimate the number of current targets and
their current states, which is a sub-problem of MTT referred to as
multi-target filtering. For the standard (point target) dynamic and
measurement models \cite{Mahler_book14}, there are multi-target conjugate
prior densities that can be used to compute or approximate the filtering
density. Multi-target conjugacy refers to a family of multi-target
distributions which is closed under both the prediction and update
steps \cite{Vo13,Williams15b,Angel18_b}. We usually consider multi-target
conjugate prior mixtures in which the number of components grows with
time. 

The Poisson multi-Bernoulli mixture (PMBM) \cite{Williams15b} is
a multi-target conjugate prior that can be written in terms of single
target densities. If the birth model is a Poisson RFS, the filtering
density is a PMBM. In this case, the Poisson part represents the targets
that have never been detected and each component of the mixture is
a global hypothesis, which has a certain weight and an associated
multi-Bernoulli density. 

A special case of a PMBM is a multi-Bernoulli mixture (MBM), which
is obtained by setting the intensity of the Poisson RFS to zero in
a PMBM. The MBM is a multi-target conjugate prior for the standard
models if the birth process is multi-Bernoulli or MBM \cite[Corollary 3]{Angel18_b}.
The resulting filter is referred to as to the MBM filter, which is
similar to the PMBM filter, but with a different processing of new
born targets. Another multi-target conjugate prior is an MBM in which
the existence probabilities of all Bernoulli components are either
0 or 1 (MBM$_{01}$). Given any MBM with probabilities of existence
between 0 and 1, it can be parameterised in MBM$_{01}$ form, but
with an exponential increase in the number of mixture components \cite{Angel18_b}.
The $\delta$-generalised labelled multi-Bernoulli ($\delta$-GLMB)
density \cite{Vo13} is a multi-target conjugate prior that is similar
in structure to an MBM$_{01}$ in which targets (Bernoulli components)
are uniquely labelled \cite[Sec. IV]{Angel18_b}. 

The main contribution of this paper is to provide a thorough description
of the MBM filter and its Gaussian implementation for linear and Gaussian
models. In the proposed implementation, we make use of Murty's algorithm
\cite{Murty68} to prune the global hypotheses, which is a common
approach in MTT \cite{Cox95,Cox96,Vo13}. We also indicate that the
MBM filter can be labelled to provide a labelled MBM filter, which
has the same filtering recursion as the MBM filter. We show simulation
results comparing the MBM filter with the PMBM and MBM$_{01}$ filters.

The rest of the paper is organised as follows. In Section \ref{sec:Problem-formulation},
we formulate the problem and introduce the relevant conjugate priors.
In Section \ref{sec:Multi-Bernoulli-mixture-filter}, we describe
the MBM filter. Section \ref{sec:Gaussian-implementation} addresses
the proposed Gaussian implementation. Simulation results are shown
in Section \ref{sec:Simulations}. Finally, conclusions are drawn
in Section \ref{sec:Conclusions}. 

\section{Problem formulation\label{sec:Problem-formulation}}

In this section, we describe the standard dynamic model, with Poisson
and multi-Bernoulli birth, and the standard measurement model. We
also explain the multi-target conjugate priors.

The set of targets at time step $k$ is denoted as $X_{k}\in\mathcal{F}\left(\mathcal{X}\right)$,
where $\mathcal{X}$ is the single target space, often $\mathcal{X}=\mathbb{R}^{n_{x}}$,
and $\mathcal{F}\left(\mathcal{X}\right)$ is the the space of all
finite subsets of $\mathcal{X}$. Given $X_{k}$, each target $x\in X_{k}$
survives to time step $k+1$ with probability $p_{S}\left(x\right)$
and moves to a new state with a transition density $g\left(\cdot\left|x\right.\right)$,
or dies with probability $1-p_{S}\left(x\right)$. Set $X_{k+1}$
is then the union of the surviving targets and new targets, which
are born independently of the rest. 

We consider two types of birth model: a Poisson RFS, which is also
called Poisson point process, with intensity $\lambda_{k}^{b}\left(\cdot\right)$,
and a multi-Bernoulli RFS, which has $n_{k}^{b}$ Bernoulli components
and the $l$-th Bernoulli component has existence probability $r_{k}^{b,l}$
and single target density $p_{k}^{b,l}\left(\cdot\right)$. The corresponding
density is \cite{Angel18_b}
\begin{align}
f_{k}^{b}\left(X_{k}\right) & =\sum_{X^{1}\uplus...\uplus X^{n_{k}^{b}}=X_{k}}\prod_{l=1}^{n_{k}^{b}}f_{k}^{b,l}\left(X^{l}\right)\label{eq:MB_birth}
\end{align}
where $\uplus$ denotes the disjoint union and the density of the
$l$-th Bernoulli component is
\begin{equation}
f_{k}^{b,l}\left(X_{k}\right)=\begin{cases}
1-r_{k}^{b,l} & X_{k}=\emptyset\\
r_{k}^{b,l}p_{k}^{b,l}\left(x\right) & X_{k}=\left\{ x\right\} \\
0 & \mathrm{otherwise}.
\end{cases}
\end{equation}
Note that in (\ref{eq:MB_birth}) the summation is taken over all
mutually disjoint (and possibly empty) sets $X^{1},...,X^{n_{k}^{b}}$
whose union is $X_{k}$. 

At time step $k$, we observe $X_{k}$ by a set $Z_{k}=\left\{ z_{k}^{1},...,z_{k}^{m_{k}}\right\} \in\mathcal{F}\left(\mathbb{R}^{n_{z}}\right)$
of measurements. Given $X_{k}$, each target state $x\in X_{k}$ is
either detected with probability $p_{D}\left(x\right)$ and generates
one measurement with density $l\left(\cdot|x\right)$, or missed with
probability $1-p_{D}\left(x\right)$. The set $Z_{k}$ is the union
of the target-generated measurements and Poisson clutter with intensity
$\kappa\left(\cdot\right)$. 

In multi-target filtering, the objective is to compute the density
of $X_{k}$ given the sequence of measurements $\left(Z_{1},...,Z_{k}\right)$.
This density can be computed recursively by the prediction and the
update steps of the filtering recursion \cite{Mahler_book14}. This
computation is aided by the use of multi-target conjugate priors

\subsection{Multi-target conjugate priors}

We first explain the PMBM conjugate prior \cite{Williams15b}. A PMBM
is the density of the union of two independent RFS: a Poisson RFS
with density $f^{p}\left(\cdot\right)$, and a multi-Bernoulli mixture
RFS with density $f^{mbm}\left(\cdot\right)$. Then, the PMBM density
is \cite{Williams15b}
\begin{align*}
f\left(X\right) & =\sum_{Y\uplus W=X}f^{p}\left(Y\right)f^{mbm}\left(W\right)\\
f^{p}\left(X\right) & =e^{-\int\lambda\left(x\right)dx}\left[\lambda\left(\cdot\right)\right]^{X}\\
f^{mbm}\left(X\right) & \propto\sum_{j}\sum_{X^{1}\uplus...\uplus X^{n}=X}\prod_{i=1}^{n}\left[w^{i,j}f^{i,j}\left(X^{i}\right)\right]
\end{align*}
where $\lambda\left(\cdot\right)$ is the intensity of the Poisson
RFS, $\propto$ stands for ``proportional to'', $j$ is an index
that goes through all the mixture components (also called global hypotheses),
$f^{i,j}\left(\cdot\right)$ is the $i$-th Bernoulli RFS in the $j$-th
global hypothesis, and $w^{i,j}$ its weight. It should be noted that
the weight $w^{j}$ of the $j$-th global hypothesis is 
\begin{align*}
w^{j} & \propto\prod_{i=1}^{n}w^{i,j}.
\end{align*}
A particular, relevant case of the PMBM is the multi-Bernoulli mixture
(MBM), which is obtained by setting $\lambda\left(\cdot\right)=0$. 

It is shown in \cite{Williams15b} that, for the Poisson birth model,
the filtering and predicted density are PMBM, which gives rise to
the PMBM filter. A corollary of this fundamental result is that the
MBM is conjugate prior if the birth model is multi-Bernoulli or MBM
\cite[Corollary 3]{Angel18_b}. In this work, we describe the MBM
filter only for multi-Bernoulli birth model as the prediction step
is simpler than for MBM birth \cite{Angel18_b}. We proceed to describe
this filter in the next section. 

\section{Multi-Bernoulli mixture filter\label{sec:Multi-Bernoulli-mixture-filter}}

The density of $X_{k'}$ with $k'\in\left\{ k,k+1\right\} $ given
the measurements up to time step $k$ is an MBM with the form
\begin{align}
f_{k'|k}\left(X_{k'}\right) & \propto\sum_{a\in\mathcal{A}_{k'|k}}\sum_{\uplus_{l=1}^{n_{k'|k}}X^{l}=X_{k'}}\prod_{i=1}^{n_{k'|k}}\left[w_{k'|k}^{i,a^{i}}f_{k'|k}^{i,a^{i}}\left(X^{i}\right)\right].\label{eq:MBM_filtering_density}
\end{align}
We proceed to explain Equation (\ref{eq:MBM_filtering_density}).
First, $n_{k'|k}$ is the number of Bernoulli components, $i$ is
an index over the Bernoulli components and a global hypothesis $a=\left(a^{1},...,a^{n_{k'|k}}\right)$
contains the single target hypotheses for all Bernoulli components.
The single target hypothesis for the $i$-th Bernoulli component is
$a^{i}=\left(t^{i},l^{i},\xi_{t^{i}:k}^{i}\right)$ where $t^{i}$
and $l^{i}$ are its birth time and birth index, see (\ref{eq:MB_birth}),
and $\xi_{t^{i}:k}^{i}=\left(\xi_{t^{i}}^{i},...,\xi_{k}^{i}\right)$
contains the corresponding data associations up to time step $k$.
In this paper, we write $\xi_{j}^{i}=0$ if the $i$-th Bernoulli
component has been misdetected at time step $j$ and $\xi_{j}^{i}=p\in\left\{ 1,...,m_{k}\right\} $
if the $i$-th Bernoulli component has been associated to the $p$-th
measurement at time step $j$. Note that, for new born Bernoulli components,
the single target hypothesis in the predicted density is only the
pair $a^{i}=\left(t^{i},l^{i}\right)$, as there has not been a data
association event for this component yet. 

In each global hypothesis $a$, a measurement $z_{k}^{j}$ can only
be assigned to one Bernoulli component, born at time step $k$ or
before, and a Bernoulli component can be assigned at most to one measurement
at each time step. All possible global hypotheses constitute the set
of global hypotheses $\mathcal{A}_{k'|k}$. Measurements left unassigned
in a global hypothesis are considered clutter under this global hypothesis.
Bernoulli components left unassigned at a particular time step in
a global hypothesis are considered misdetected. 

The density $f_{k'|k}^{i,a^{i}}\left(\cdot\right)$ of the $i$-th
Bernoulli component with single target hypothesis $a^{i}$ is written
as
\begin{align}
f_{k'|k}^{i,a^{i}}\left(X\right) & =\begin{cases}
1-r_{k'|k}^{i,a^{i}} & X=\emptyset\\
r_{k'|k}^{i,a^{i}}p_{k'|k}^{i,a^{i}}\left(x\right) & X=\left\{ x\right\} \\
0 & \mathrm{otherwise}
\end{cases}\label{eq:Bernoulli_density_filter}
\end{align}
and has an associated weight $w_{k'|k}^{i,a^{i}}$. 

In the rest of this section, we explain the prediction and update
steps to recursively compute (\ref{eq:MBM_filtering_density}) in
Sections \ref{subsec:Prediction_MBM} and \ref{subsec:Update-MBM}.
A discussion of the recursion is given in Section \ref{subsec:Discussion}.
We will use the following notation for the inner product of two functions
$h\left(\cdot\right)$ and $g\left(\cdot\right)$
\begin{align*}
\left\langle h,g\right\rangle  & =\int h\left(x\right)g\left(x\right)dx.
\end{align*}

\subsection{Prediction\label{subsec:Prediction_MBM}}

We consider that the filtering density at time step $k$ is an MBM
of the form (\ref{eq:MBM_filtering_density}) with $k'=k$. Then,
the predicted density has the same number of global hypotheses as
the filtering density but with $n_{k+1|k}=n_{k|k}+n_{k+1}^{b}$ Bernoulli
components. That is, each global hypothesis is augmented with the
Bernoulli components that represent new born targets. For the surviving
Bernoulli components, $i\in\left\{ 1,...,n_{k|k}\right\} $, the parameters
are
\begin{align}
w_{k+1|k}^{i,a^{i}} & =w_{k|k}^{i,a^{i}}\label{eq:predicted_first}\\
r_{k+1|k}^{i,a^{i}} & =r_{k|k}^{i,a^{i}}\left\langle p_{S},p_{k|k}^{i,a^{i}}\right\rangle \label{eq:predicted_existence}\\
p_{k+1|k}^{i,a^{i}}\left(x\right) & =\frac{\int g\left(x\left|y\right.\right)p_{S}\left(y\right)p_{k|k}^{i,a^{i}}\left(y\right)dy}{\left\langle p_{k|k}^{i,a^{i}},p_{S}\right\rangle }.\label{eq:predicted_density}
\end{align}

For Bernoulli components of new born targets, $i\in\left\{ n_{k|k}+1,...,n_{k+1|k}\right\} $,
the parameters are
\begin{align}
a^{i} & =\left(k+1,i-n_{k|k}\right)\\
w_{k+1|k}^{i,a^{i}} & =1\\
r_{k+1|k}^{i,a^{i}} & =r_{k+1}^{b,i-n_{k|k}}\\
p_{k+1|k}^{i,a^{i}}\left(x\right) & =p_{k+1}^{b,i-n_{k|k}}\left(x\right).\label{eq:predicted_last}
\end{align}

\subsection{Update\label{subsec:Update-MBM}}

We recall that the set of measurements at time step $k$ is denoted
as $Z_{k}=\left\{ z_{k}^{1},...,z_{k}^{m_{k}}\right\} $. The number
of Bernoulli components does not change in the update so $n_{k|k}=n_{k|k-1}$.
The update of the $i$-th Bernoulli component is as follows. We go
through all single target hypotheses and create misdetection and measurement
associated hypotheses. In this section, we use $\left(a^{i},p\right)$,
with $p\in\left\{ 0,...,m_{k}\right\} $ to append $p$ to the single
target hypothesis $a^{i}$. For a single target hypothesis $a^{i}$
at the previous time step, the misdetection single target hypothesis
is characterised by 
\begin{align}
w_{k|k}^{i,\left(a^{i},0\right)} & =w_{k|k-1}^{i,a^{i}}\nonumber \\
 & \,\times\left(1-r_{k|k-1}^{i,a^{i}}+r_{k|k-1}^{i,a^{i}}\left\langle p_{k|k-1}^{i,a^{i}},1-p_{D}\right\rangle \right)\label{eq:update_mis_first}\\
r_{k|k}^{i,\left(a^{i},0\right)} & =\frac{r_{k|k-1}^{i,a^{i}}\left\langle p_{k|k-1}^{i,a^{i}},1-p_{D}\right\rangle }{1-r_{k|k-1}^{i,a^{i}}+r_{k|k-1}^{i,a^{i}}\left\langle p_{k|k-1}^{i,a^{i}},1-p_{D}\right\rangle }\\
p_{k|k}^{i,\left(a^{i},0\right)}\left(x\right) & =\frac{\left(1-p_{D}\left(x\right)\right)p_{k|k-1}^{i,a^{i}}\left(x\right)}{\left\langle p_{k|k-1}^{i,a^{i}},1-p_{D}\right\rangle }.\label{eq:update_mist_last}
\end{align}
The corresponding updated single target hypothesis with measurement
$z_{k}^{j}$ is characterised by
\begin{align}
w_{k|k}^{i,\left(a^{i},j\right)} & =\frac{w_{k|k-1}^{i,a^{i}}r_{k|k-1}^{i,a^{i}}\left\langle p_{k|k-1}^{i,a^{i}},p_{D}l\left(z_{k}^{j}|\cdot\right)\right\rangle }{\kappa\left(z_{k}^{j}\right)}\label{eq:update_weight_detection}\\
r_{k|k}^{i,\left(a^{i},j\right)} & =1\\
p_{k|k}^{i,\left(a^{i},j\right)}\left(x\right) & =\frac{p_{D}\left(x\right)l\left(z_{k}^{j}|x\right)p_{k|k-1}^{i,a^{i}}\left(x\right)}{\left\langle p_{k|k-1}^{i,a^{i}},p_{D}l\left(z_{k}^{j}|\cdot\right)\right\rangle }.\label{eq:update_det_last}
\end{align}

Once we have formed the single target hypotheses for all Bernoulli
components, a previous global hypothesis generates new global hypotheses
that correspond to the possible associations of measurements to Bernoulli
components, such that one measurement can be assigned to at most one
Bernoulli component and one Bernoulli component can only be assigned
to at most one measurement. 

The MBM filter update is equivalent to the PMBM update in \cite[Thm. 3]{Williams15b}
by setting the intensity of the Poisson component of the PMBM to zero
\cite{Angel18_b}. Nevertheless, there is a difference in how the
update has been written in this section and in \cite[Thm. 3]{Williams15b}.
The update in \cite[Thm. 3]{Williams15b} creates a new Bernoulli
component for each measurement, which represents a potential target,
as it can be clutter or a real target. A potential target created
by a given measurement exists in global hypotheses in which this measurement
is not assigned to previously existing Bernoulli components. If the
intensity of the Poisson process is zero, a measurement that has not
been assigned to a previously existing Bernoulli component is not
a potential target, but clutter with probability one. This sets the
probability of existence of the corresponding Bernoulli to zero. While
this approach is correct, the Bernoulli components created in this
fashion will always have a probability of existence equal to zero,
so it is more suitable in the MBM filter not to create these components.
In this case, the weights of the detected hypotheses must be adjusted
to leave the weights of the global hypotheses unaltered. To this end,
denominator $\kappa\left(z_{k}^{j}\right)$ is included in \cite[Eq. (49)]{Williams15b}
to yield (\ref{eq:update_weight_detection}).  

\subsection{Discussion\label{subsec:Discussion}}

As pointed out in \cite[Sec. IV]{Angel18_b}, with MB birth, one can
obtain the corresponding labelled MBM filter, as a particular case
of the above MBM filter. The MBM filtering recursion has been provided
for a general single target state $x$. This state is general enough
to accommodate a label \cite{Angel09,Angel13,Vo13,Vo14,Aoki16}, which
can be written as $x=\left(x',\ell\right)$ where $\ell$ is the label,
which are uniquely assigned to each Bernoulli component and $x'$
is the rest of the target state. The uniqueness of the labels is achieved
by considering the particular case in which 
\begin{itemize}
\item The density $p_{k}^{b,l}\left(\cdot\right)$ in MB birth (\ref{eq:MB_birth})
is $p_{k}^{b,l}\left(\left(x',\ell\right)\right)=p_{k}^{b,l}\left(x'\right)\delta\left[\ell-\left(k,l\right)\right]$,
This ensures that each Bernoulli component is uniquely labelled upon
birth.
\item The single target transition density $g\left(\cdot\left|\cdot\right.\right)$
is $g\left(\left(x',\ell_{x}\right)\left|\left(y',\ell_{y}\right)\right.\right)=\delta\left[\ell_{x}-\ell_{y}\right]g\left(x'\left|y'\right.\right)$.
This ensures that each Bernoulli component does not change its (unique)
label.
\end{itemize}
Both unlabelled and labelled MBM filters are implemented in the same
way as they follow the same recursion. In both filters, labels are
part of the single target hypotheses, so they always belong to the
metadata of the filters. This equivalence in the filtering recursion
between unlabelled and labelled approaches also holds in MTT using
sets of trajectories \cite[Sec. IV.A]{Angel15_prov}. From the MBM
filtering recursion, one can also obtain the MBM$_{01}$ filtering
recursion. The MBM$_{01}$ filter is analogous to the MBM filter with
the additional step that after prediction step the resulting MBM density
is parameterised in its MBM$_{01}$ form \cite[Sec. IV.A]{Angel18_b}.
This operation entails an exponential increase in the number of global
hypotheses in general settings, so the MBM filter is preferable over
the MBM$_{01}$ filter.

It is also relevant to discuss the choice of birth model, either Poisson
or multi-Bernoulli. A multi-Bernoulli birth can be suitable if one
is certain that a known maximum of targets will enter the area of
interest and the targets appear around some known locations. In this
case, one can put a number of Bernoulli components (usually one) in
each location to account for possible births. A practical example
can be the tracking of people in a room with several doors where only
one person can pass each door at a time. Overlapping Bernoulli components
can also be used to cover potential births in large areas, but in
this case, one must be certain that the number of appearing targets
does not exceed the predefined number of Bernoulli components. Adding
Bernoulli components to the birth model increases the computational
burden. A problem with the multi-Bernoulli birth model occurs when
there is a modelling error and the number of new born targets that
are detected is higher than the number of birth components. In this
case, the filter will not be able to estimate a state for each target
at the time of the first detection so there will be missed target
errors. 

On the contrary, the Poisson birth model does not set a maximum to
the number of targets. It can model target births at known points
sources and also cover large areas of potential births, representing
the information efficiently using its intensity. Therefore, Poisson
models seem more suitable in radar surveillance applications in broad
areas \cite{Williams15b} and robotic applications \cite{Cament18}.
Also, when prior birth information is vague, it is more sensible to
initiate Bernoulli components based on the measurements, as in the
PMBM filter. 

It should also be noted that a Bernoulli RFS with low existence probability
can be approximated very accurately by a Poisson RFS \cite{Williams12},
without the constraint of setting a maximum number of targets. Therefore,
if the targets are born at known locations and the constraint on the
maximum number of appearing targets of multi-Bernoulli birth is met,
one should not expect a large difference between the models. 

Finally, we would like to mention that one benefit of the Poisson
part in the PMBM filter, which is missing in the MBM filter, is that
it allows for the use of recycling \cite{Williams12}. In this technique,
Bernoulli components removed in pruning are merged into the Poisson
part rather than being completely discarded, which can be used to
lower computational cost without sacrificing performance \cite{Xia17}.

\section{Gaussian implementation with Murty's algorithm\label{sec:Gaussian-implementation}}

The Gaussian implementation is obtained when there are constant probabilities
$p_{S}$ and $p_{D}$ of survival and detection and Gaussian/linear
models 
\begin{itemize}
\item $g\left(x\left|y\right.\right)=\mathcal{N}\left(x;Fy,Q\right)$,
\item $l\left(z|x\right)=\mathcal{N}\left(z;Hx,R\right)$, 
\item $p_{k}^{b,l}\left(x\right)=\mathcal{N}\left(x;\overline{x}_{k}^{b,l},P_{k}^{b,l}\right)$,
\end{itemize}
where $\mathcal{N}\left(x;\overline{x},P\right)$ denotes a Gaussian
density with mean $\overline{x}$ and covariance matrix $P$ evaluated
at $x$. In this case, the predicted and filtering densities are MBM
of the form (\ref{eq:MBM_filtering_density}) with Gaussian single-target
densities
\begin{align*}
p_{k'|k}^{i,a^{i}}\left(x\right) & =\mathcal{N}\left(x;\overline{x}_{k'|k}^{i,a^{i}},P_{k'|k}^{i,a^{i}}\right)
\end{align*}

\subsection{Prediction\label{subsec:Prediction_GM}}

The prediction is given by Equations (\ref{eq:predicted_first})-(\ref{eq:predicted_last}).
For the linear/Gaussian models, these expressions for $i\in\left\{ 1,...,n_{k|k}\right\} $
can be written as
\begin{align*}
w_{k+1|k}^{i,a^{i}} & =w_{k|k}^{i,a^{i}}\\
r_{k+1|k}^{i,a^{i}} & =r_{k|k}^{i,a^{i}}p_{S}\\
p_{k+1|k}^{i,a^{i}}\left(x\right) & =\mathcal{N}\left(x;\overline{x}_{k+1|k}^{i,a^{i}},P_{k+1|k}^{i,a^{i}}\right)
\end{align*}
where
\begin{align*}
\overline{x}_{k+1|k}^{i,a^{i}} & =F\overline{x}_{k|k}^{i,a^{i}}\\
P_{k+1|k}^{i,a^{i}} & =FP_{k|k}^{i,a^{i}}F^{T}+Q.
\end{align*}

\subsection{Update\label{subsec:Update_GM}}

The update of single target hypotheses corresponding to misdetections
is given by (\ref{eq:update_mis_first})-(\ref{eq:update_mist_last}),
which simplify as
\begin{align}
w_{k|k}^{i,\left(a^{i},0\right)} & =w_{k|k-1}^{i,a^{i}}\left(1-r_{k|k-1}^{i,a^{i}}+r_{k|k-1}^{i,a^{i}}\left(1-p_{D}\right)\right)\label{eq:misdetection_GM1}\\
r_{k|k}^{i,\left(a^{i},0\right)} & =\frac{r_{k|k-1}^{i,a^{i}}\left(1-p_{D}\right)}{1-r_{k|k-1}^{i,a^{i}}+r_{k|k-1}^{i,a^{i}}\left(1-p_{D}\right)}\\
p_{k|k}^{i,\left(a^{i},0\right)}\left(x\right) & =p_{k|k-1}^{i,a^{i}}\left(x\right).\label{eq:misdetection_GM_last}
\end{align}
The updated single target hypothesis with measurement $z_{k}^{j}$,
which is given by (\ref{eq:update_weight_detection})-(\ref{eq:update_det_last}),
is characterised by
\begin{align}
w_{k|k}^{i,\left(a^{i},j\right)} & =\frac{w_{k|k-1}^{i,a^{i}}r_{k|k-1}^{i,a^{i}}p_{D}\mathcal{N}\left(z_{k}^{j};H\overline{x}_{k|k-1}^{i,a^{i}},S_{k|k-1}^{i,a^{i}}\right)}{\kappa\left(z_{k}^{j}\right)}\label{eq:detection_GM1}\\
r_{k|k}^{i,\left(a^{i},j\right)} & =1\\
p_{k|k}^{i,\left(a^{i},j\right)}\left(x\right) & =\mathcal{N}\left(x;\overline{x}_{k|k}^{i,\left(a^{i},j\right)},P_{k|k}^{i,\left(a^{i},j\right)}\right)
\end{align}
where
\begin{align}
\overline{x}_{k|k}^{i,\left(a^{i},j\right)} & =\overline{x}_{k|k-1}^{i,a^{i}}+P_{k|k-1}^{i,a^{i}}H^{T}\left(S_{k|k-1}^{i,a^{i}}\right)^{-1}\left(z_{k}^{j}-H\overline{x}_{k|k-1}^{i,a^{i}}\right)\\
P_{k|k}^{i,\left(a^{i},j\right)} & =P_{k|k-1}^{i,a^{i}}-P_{k|k-1}^{i,a^{i}}H^{T}\left(S_{k|k-1}^{i,a^{i}}\right)^{-1}HP_{k|k-1}^{i,a^{i}}\\
S_{k|k-1}^{i,a^{i}} & =HP_{k|k-1}^{i,a^{i}}H^{T}+R.\label{eq:detection_GM_last}
\end{align}

\subsection{Practical implementation}

The MBM filtering recursion explained above cannot be carried out
without approximations in practice, due to the ever increasing number
of hypotheses and Bernoulli components. To this end, we perform pruning
of global hypotheses and Bernoulli components. In order to explain
how we perform pruning in the proposed implementation, it is convenient
to write the filtering/predicted density in (\ref{eq:MBM_filtering_density})
as
\begin{align}
f_{k'|k}\left(X_{k'}\right) & =\sum_{a\in\mathcal{A}_{k'|k}}w_{k'|k}^{a}\sum_{\uplus_{l=1}^{n_{k'|k}}X^{l}=X_{k'}}\prod_{i=1}^{n_{k'|k}}f_{k'|k}^{i,a^{i}}\left(X^{i}\right)\label{eq:MBM_filtering_density_explicit_weight}
\end{align}
where the weight of global hypothesis $a$ is
\begin{align*}
w_{k'|k}^{a} & \propto\prod_{i=1}^{n_{k'|k}}w_{k'|k}^{i,a^{i}}.
\end{align*}

In this representation, the information of the filtering/predicted
densities, see (\ref{eq:MBM_filtering_density_explicit_weight}),
is stored as
\begin{itemize}
\item $n_{k'|k}$ Bernoulli components. The $i$-th Bernoulli component
has
\begin{itemize}
\item Bernoulli densities $f_{k'|k}^{i,a^{i}}\left(\cdot\right)$ for all
the single target hypotheses for the $i$-th Bernoulli component.
Each Bernoulli density is parameterised by $r_{k'|k}^{i,a^{i}}$ and
$p_{k'|k}^{i,a^{i}}\left(\cdot\right)$, see (\ref{eq:Bernoulli_density_filter}).
\end{itemize}
\item Global hypotheses. Each global hypothesis consists of its weight,
and a list of $n_{k'|k}$ pointers that indicate the single target
hypothesis for each Bernoulli component that belongs to this global
hypothesis. 
\end{itemize}
Pruning the global hypotheses consists of approximating some of the
weights $w_{k'|k}^{a}$ as zero, followed by weight renormalisation,
so that these global hypotheses are removed and do not have to be
propagated through the filtering recursion. It should be noted that,
clearly, setting some of the global hypothesis weights to zero does
not affect the symmetry of the density w.r.t. the elements of set
argument $X_{k'}=\left\{ x_{1},...,x_{n}\right\} $, as each of the
terms in the mixture is a multi-Bernoulli density, which is symmetric. 

Pruning is performed at two stages, at the update step and after target
state estimation. We proceed to describe both.

\subsubsection{Pruning at the update step}

At the update step, one can perform pruning before enumerating all
newly generated global hypotheses. The first technique to limit the
number of new global hypotheses is ellipsoidal gating \cite{Kurien_inbook90}.
In order to so, when we go through each Bernoulli component to create
a new single target hypothesis with measurement $z_{k}^{j}$, we evaluate
\begin{align}
\left(z_{k}^{j}-H\overline{x}_{k|k-1}^{i,a^{i}}\right)^{T}\left(S_{k|k-1}^{i,a^{i}}\right)^{-1}\left(z_{k}^{j}-H\overline{x}_{k|k-1}^{i,a^{i}}\right).\label{eq:gating}
\end{align}
If (\ref{eq:gating}) is greater than a predefined threshold $\Gamma_{g}$,
this updated single target hypothesis is not created. 

Given a global hypothesis $a$ at the previous time step, in theory,
we must go through all possible data association hypotheses that give
rise to the updated global hypotheses. Nevertheless, we can perform
pruning and select the $k_{u}$ new global hypotheses with highest
weight for a given global hypothesis $a$ without evaluating all the
newly generated global hypotheses. To this end, we use Murty's algorithm
\cite{Murty68}, which requires an algorithm to solve assignment problems.
In our implementations, we have used the Hungarian algorithm \cite{Kuhn55}. 

The cost matrix $C$ of the assignment problem for global hypothesis
$a$ is of dimensions $n_{k|k-1}\times m_{k}$. The $i,j$ component
of $C$ is 
\begin{align}
C_{i,j} & =-\ln\left(\frac{w_{k|k}^{i,\left(a^{i},j\right)}}{w_{k|k}^{i,\left(a^{i},0\right)}}\right)\nonumber \\
 & =-\ln\left(\frac{r_{k|k-1}^{i,a^{i}}p_{D}\mathcal{N}\left(z_{k}^{j};H\overline{x}_{k|k-1}^{i,a^{i}},S_{k|k-1}^{i,a^{i}}\right)}{\kappa\left(z_{k}^{j}\right)\left(1-r_{k|k-1}^{i,a^{i}}+r_{k|k-1}^{i,a^{i}}\left(1-p_{D}\right)\right)}\right).\label{eq:cost_matrix}
\end{align}
We would like to remark that, for the new single target hypotheses
that did not pass the gating threshold, one sets $C_{i,j}=-\infty$,
which comes from the approximation $\mathcal{N}\left(z_{k}^{j};H\overline{x}_{k|k-1}^{i,a^{i}},S_{k|k-1}^{i,a^{i}}\right)\simeq0$.
This ensures that the chosen global hypotheses do not contain single
target hypotheses that have not passed the gating threshold. 

A new global hypothesis (assignment) can be represented by a matrix
$S$, whose entries are 0 or 1, with every row and column summing
to either 1 or 0. $S_{i,j}=1$ if and only if the $j$-th measurement
is associated with the $i$-th Bernoulli component. From a previous
global hypothesis $a$, the weight of the new global hypothesis parameterised
by $S$ is \cite{Vo14} proportional to
\begin{align*}
 & w_{k|k-1}^{a}\exp\left(-\mathrm{tr}\left(S^{T}C\right)\right)\prod_{i=1}^{n_{k|k}}\left(1-r_{k|k-1}^{i,a^{i}}+r_{k|k-1}^{i,a^{i}}\left(1-p_{D}\right)\right).
\end{align*}
Therefore, the $k_{u}$ global hypotheses with highest weight can
be found by solving the $k_{u}$ assignment matrices that minimise
$\mathrm{tr}\left(S^{T}C\right)$, for which we use Murty's algorithm.
We select $k_{u}=\left\lceil N_{h}\cdot w_{k|k-1}^{a}\right\rceil $,
where $N_{h}$ is the maximum number of global hypotheses as in \cite{Vo14,Angel18_b}. 

It is relevant to notice that the costs of the assignment problem
for the $\delta$-GLMB filter \cite[Eq. (24)]{Vo14} are equivalent
to the costs of the MBM filter if the existence probabilities are
equal to one, $r_{k|k-1}^{i,a^{i}}=1$. This is due to the fact that
each global hypothesis in the the $\delta$-GLMB considers targets
with deterministic target existence, similar to the MBM$_{01}$ filter,
rather than probabilistic, as in the MBM/PMBM filters. 

\subsubsection{Pruning after estimation\label{subsec:Pruning-after-estimation}}

After multi-target state estimation, we perform pruning of Bernoulli
components and global hypotheses following these three steps
\begin{enumerate}
\item Keep the global hypotheses that have the $N_{h}$ highest weights,
and whose weight is higher than a threshold.
\item Remove the single target hypotheses of the Bernoulli components that
do not take part in any of the considered global hypotheses. 
\item Remove the Bernoulli components whose existence is lower than a threshold
$\Gamma_{b}$ for all its single target hypotheses.
\end{enumerate}
As a result of the previous pruning operations, there can be global
hypotheses that have the same single target hypotheses for all Bernoulli
components. As these global hypotheses are alike, they are merged
into one, whose weight is the sum of the merged global hypotheses,
to save computational resources. 

\subsection{Estimation\label{subsec:Estimation}}

The computationally efficient estimators of the PMBM filter explained
in \cite[Sec. VI]{Angel18_b}, can be directly applied to the MBM
filter, as they do not take into consideration the Poisson component.
Finally, the pseudocode of one prediction and update are given in
Algorithm \ref{alg:MBM_pseudocode}. 

\selectlanguage{british}%
\begin{algorithm}
\selectlanguage{english}%
\caption{\foreignlanguage{british}{\label{alg:MBM_pseudocode}\foreignlanguage{english}{Prediction and
update steps for the MBM filter}}}

{\fontsize{9}{9}\selectfont

\begin{algorithmic}     

\State - Perform prediction, see Section \ref{subsec:Prediction_GM}.

\State $\phantom{}$ \foreignlanguage{british}{\Comment{Update}}

\selectlanguage{british}%
\For{$i=1$ to $n_{k|k-1}$} \Comment{Go through all Bernoulli components}

\ForAll{$a^{i}$}\Comment{Go through all its single target hypotheses}

\selectlanguage{english}%
\State - Create new misdetection hypothesis, see (\ref{eq:misdetection_GM1})-(\ref{eq:misdetection_GM_last}). 

\selectlanguage{british}%
\For{$j=1$ to $m_{k}$} \Comment{Go through the measurements}

\selectlanguage{english}%
\State - If $z_{k}^{j}$ satisfies gating condition, see (\ref{eq:gating}),
create a new detection hypothesis using (\ref{eq:detection_GM1})-(\ref{eq:detection_GM_last}). 

\selectlanguage{british}%
\EndFor

\EndFor

\EndFor

\ForAll{$a$} \Comment{Go through all previous global hypotheses}

\selectlanguage{english}%
\State - Create cost matrix with elements in (\ref{eq:cost_matrix}).

\State - Run Murty's algorithm to select $k_{u}=\left\lceil N_{h}\cdot w_{k|k-1}^{a}\right\rceil $
new global hypotheses. 

\selectlanguage{british}%
\EndFor

\selectlanguage{english}%
\State - Estimate target states, see Section \ref{subsec:Estimation}.

\State $\phantom{}$ \foreignlanguage{british}{\Comment{Pruning}}

\State - Prune global hypotheses and Bernoulli components, see Section
\ref{subsec:Pruning-after-estimation}.

\selectlanguage{british}%
\end{algorithmic}

}
\end{algorithm}
\selectlanguage{english}%

\section{Simulations\label{sec:Simulations}}

In this section, we evaluate the performance of the MBM filter against
other algorithms in the literature. We evaluate the algorithms using
the generalised optimal sub-pattern assignment (GOSPA) metric\footnote{Matlab code of the GOSPA metric and its decomposition can be found
in https://github.com/abusajana/GOSPA} \cite{Rahmathullah17} with $\alpha=2$, as it is only for this value
of $\alpha$ that the metric decomposes into localisation errors for
properly detected targets and costs for missed and false targets. 

\subsection{GOSPA metric and its decomposition}

Given $c>0$, $1\leq p<\infty$, a metric $d\left(\cdot,\cdot\right)$
in the single target space, the ground truth set $X_{k}=\left\{ x_{k}^{1},...,x_{k}^{\left|X_{k}\right|}\right\} $
and its estimate $\hat{X}_{k}=\left\{ x_{k}^{1},...,x_{k}^{\left|\hat{X}_{k}\right|}\right\} $,
the GOSPA metric for $\alpha=2$ (not for other values) can be written
as \cite[Prop. 1]{Rahmathullah17}
\begin{align*}
 & d_{p}^{\left(c,2\right)}\left(X_{k},\hat{X}_{k}\right)\\
 & =\min_{\gamma\in\Gamma}\left(\sum_{\left(i,j\right)\in\gamma}d^{p}\left(x_{k}^{i},\hat{x}_{k}^{j}\right)+\frac{c^{p}}{2}\left(\left|X_{k}\right|+\left|\hat{X}_{k}\right|-2\left|\gamma\right|\right)\right)^{\frac{1}{p}}
\end{align*}
where $\gamma$ is an assignment set between $\left\{ 1,...,\left|X_{k}\right|\right\} $
and $\left\{ 1,...,\left|\hat{X}_{k}\right|\right\} $, which meets
$\gamma\subseteq\left\{ 1,...,\left|X_{k}\right|\right\} \times\left\{ 1,...,\left|\hat{X}_{k}\right|\right\} $,
$\left(i,j\right),\left(i,j'\right)\in\gamma\rightarrow j=j'$, and
$\left(i,j\right),\left(i',j\right)\in\gamma\rightarrow i=i'$. The
last two properties ensure that every $i$ and $j$ gets at most one
assignment. The set $\Gamma$ denotes the set of all possible $\gamma$.
Also, note that there is no cut-off parameter for $d\left(\cdot,\cdot\right)$,
which is required when the metric is written in terms of permutations
\cite[Eq. (1)]{Rahmathullah17}. 

Let $\gamma^{\star}$ denote the optimal assignment in the GOSPA metric.
Then, the GOSPA metric can be decomposed as 
\begin{align*}
d_{p}^{\left(c,2\right)}\left(X_{k},\hat{X}_{k}\right) & =\left[c_{l}^{p}\left(X_{k},\hat{X}_{k},\gamma^{\star}\right)+c_{m}^{p}\left(\gamma^{\star}\right)+c_{f}^{p}\left(\gamma^{\star}\right)\right]^{\frac{1}{p}}
\end{align*}
where $c_{l}^{p}\left(\cdot\right)$ is the localisation cost for
properly detected targets to the $p$-th power, $c_{m}^{p}\left(\cdot\right)$
is the missed target cost to the $p$-th power and $c_{f}^{p}\left(\cdot\right)$
is the false target cost to the $p$-th power. These costs have the
expressions
\begin{align*}
c_{l}^{p}\left(X_{k},\hat{X}_{k},\gamma^{\star}\right) & =\sum_{\left(i,j\right)\in\gamma^{\star}}d^{p}\left(x_{k}^{i},\hat{x}_{k}^{j}\right)\\
c_{m}^{p}\left(\gamma^{\star}\right) & =\frac{c^{p}}{2}\left(\left|X_{k}\right|-\left|\gamma^{\star}\right|\right)\\
c_{f}^{p}\left(\gamma^{\star}\right) & =\frac{c^{p}}{2}\left(\left|\hat{X}_{k}\right|-\left|\gamma^{\star}\right|\right)
\end{align*}
where $\left|X_{k}\right|-\left|\gamma^{\star}\right|$ and $\left|\hat{X}_{k}\right|-\left|\gamma^{\star}\right|$
are the number of missed and false targets, respectively.

\subsection{Comparison}

We show simulation results that compare the MBM\footnote{Matlab codes of the PMBM and MBM filters can be found in https://github.com/Agarciafernandez/MTT}
filter with the PMBM filter \cite{Angel18_b}, and the $\text{MBM}_{01}$
filter, which is similar to the $\delta$-GLMB filter. For the implementation
of the $\text{MBM}_{01}$ filter, we consider the joint prediction
and update formulation of the assignment problem. This idea was first
proposed in \cite{Correa15}, and then used in the $\delta$-GLMB
filter in \cite{Vo17}. Murty's algorithm is used in all the compared
filters to obtain global hypotheses with the highest weights. All
units in this section are given in the international system. 

Target states consist of 2D position and velocity $[p_{x},v_{x},p_{y},v_{y}]^{T}$
with dynamics characterised by
\[
F=I_{2}\otimes\begin{bmatrix}1 & T\\
0 & 1
\end{bmatrix},\,Q=qI_{2}\otimes\begin{bmatrix}T^{3}/3 & T^{2}/2\\
T^{2}/2 & T
\end{bmatrix},
\]
where $\otimes$ is the Kronecker product, $q=0.01$, and the sampling
time $T=1$. We also consider $p_{S}=0.99$. 

We measure the position of the targets with the model
\[
H=I_{2}\otimes\begin{bmatrix}1 & 0\end{bmatrix},\,R=I_{2}.
\]
The Poisson clutter is uniformly distributed in the region of interest
$A=[0,300]\times[0,300]$ . Therefore, $\kappa\left(z\right)=\lambda_{c}\cdot u_{A}\left(z\right)$
where $u_{A}\left(z\right)$ is a uniform density and $\lambda_{c}=10$,
which implies 10 expected false alarms per time step. The probability
of detection is $p_{D}=0.9$ .

The filters consider that there are no targets at time 0. For all
the compared filters, global hypotheses with weight smaller than $10^{-5}$
are pruned, and the number of global hypotheses is capped at $N_{h}$.
We will analyse performance for $N_{h}\in\left\{ 100,200,300,400,500\right\} $.
For the PMBM filter, we also remove mixture components in the Poisson
point process intensity with weights smaller than $\Gamma_{b}=10^{-5}$,
without recycling. In addition, for the PMBM filter and the MBM filter,
Bernoulli components with existence probability smaller than $10^{-3}$
are pruned. We also use ellipsoidal gating with threshold $\Gamma_{g}=20$.

\begin{figure}[t]
\centering \includegraphics[width=0.8\linewidth]{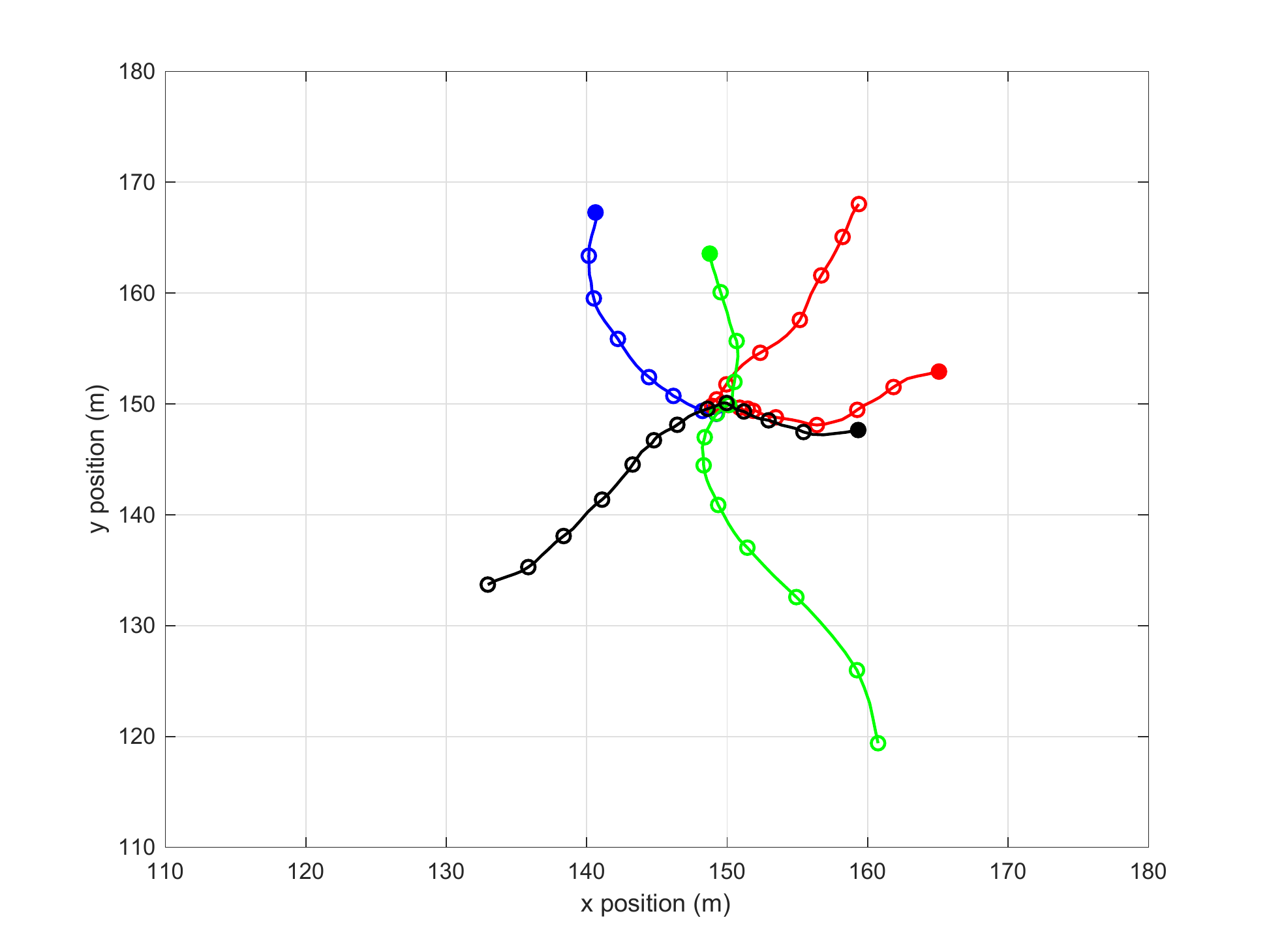}
\caption{True target trajectories of the four considered targets. The blue
one and the red one are born at time step 1, whereas the green one
and the black one are born at time step 21. The only target that dies
during the simulation is the blue target, which dies at time step
40, when all targets are in close proximity. Targets positions every
10 time steps are marked with a circle, and their initial positions
with a filled circle.}
\label{fig:groundtruth} 
\end{figure}

We consider 81 time steps and the true target trajectories in Figure
\ref{fig:groundtruth}. For each trajectory, we initiate the midpoint
(state at time step 41) from a Gaussian with mean $[150,0,150,0]^{T}$
and covariance matrix $0.01I_{4}$. The rest of the trajectory is
generated by running forward and backward dynamics. This scenario
is challenging due to the high number of targets in close proximity,
and the fact that one target dies when targets are in close proximity.

For the PMBM filter, the Poisson birth intensity has the form $\lambda_{k}^{b}(x)=\sum_{l=1}^{n_{p}}\lambda_{k}^{b,l}\mathcal{N}(x;\bar{x}_{k}^{b,l},P_{k}^{b,l})$.
For the MBM filter and the $\text{MBM}_{01}$ filter, the $l$th Bernoulli
component in the MB birth density has existence probability $r_{k}^{b,l}$
and single target state density $\mathcal{N}(x;\bar{x}_{k}^{b,l},P_{k}^{b,l})$,
where $1\leq l\leq n_{b}$. To evaluate the estimation performance
of the compared filters under different birth parameter settings,
three scenarios were simulated. In the first scenario, we consider
a case where targets can be born at several known locations with low
uncertainty, as in \cite{Vo13}. We set $n_{p}=n_{b}=4$, $\lambda_{k}^{b,l}=r_{k}^{b,l}=0.01$,
and $P_{k}^{b,l}=\text{diag}([3,1,3,1])^{2}$. The mean of the Gaussian
components are $\bar{x}_{k}^{b,1}=[140,0,170,0]^{T}$, $\bar{x}_{k}^{b,2}=[165,0,155,0]^{T}$,
$\bar{x}_{k}^{b,3}=[150,0,160,0]^{T}$ and $\bar{x}_{k}^{b,4}=[160,0,150,0]$,
respectively. In the second scenario, we consider a case where targets
can be born in a broad area that covers the region of interest, as
in \cite{Williams15b}. We set $n_{p}=1$, $\lambda_{k}^{b,1}=0.04$,
$n_{b}=2$, $r_{k}^{b,l}=0.02$, $\bar{x}_{k}^{b,l}=[100,0,100,0]^{T}$
and $P_{k}^{b,l}=\text{diag}([150,1,150,1])^{2}$. In the third scenario,
we consider a case where targets do not generate measurements until
time step 10, which means that $p_{D}=0$ for the first 10 time steps
and then $p_{D}=0.9$. The value of $p_{D}$ at each time step is
known by the filters. The birth parameter settings are the same as
in the second scenario. It should be noted that in all scenarios the
multi-Bernoulli and Poisson birth model have the same intensity (probability
hypothesis density) \cite[Eq. (4.129)]{Mahler_book14}. This implies
that birth models are as close as possible from a Kullback-Leibler
divergence perspective. 

\begin{figure}[!]
\begin{centering}
\subfloat[Scenario 1]{\begin{centering}
\includegraphics[width=1\linewidth]{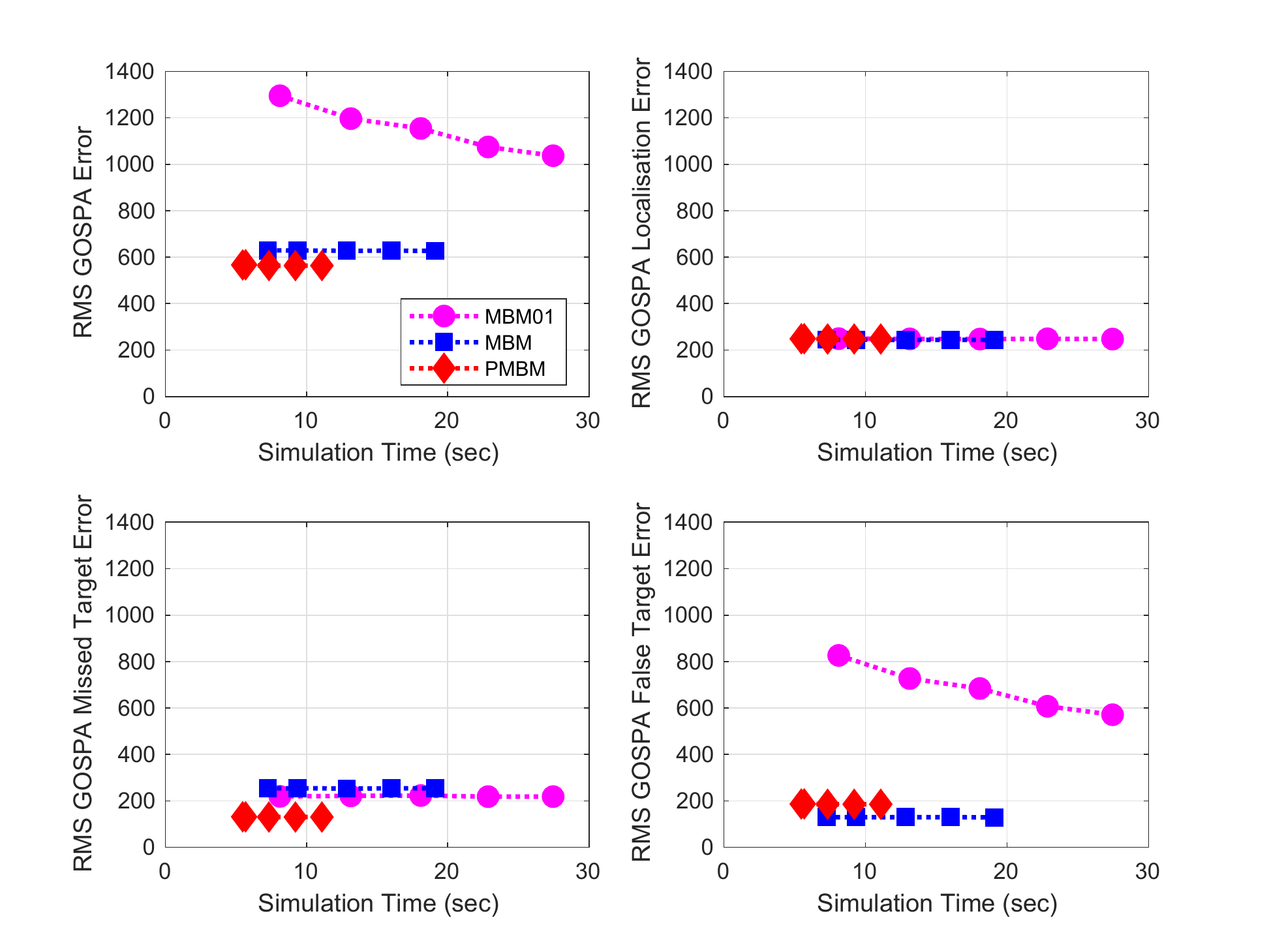}
\par\end{centering}
}
\par\end{centering}
\begin{centering}
\subfloat[Scenario 2]{\begin{centering}
\includegraphics[width=1\linewidth]{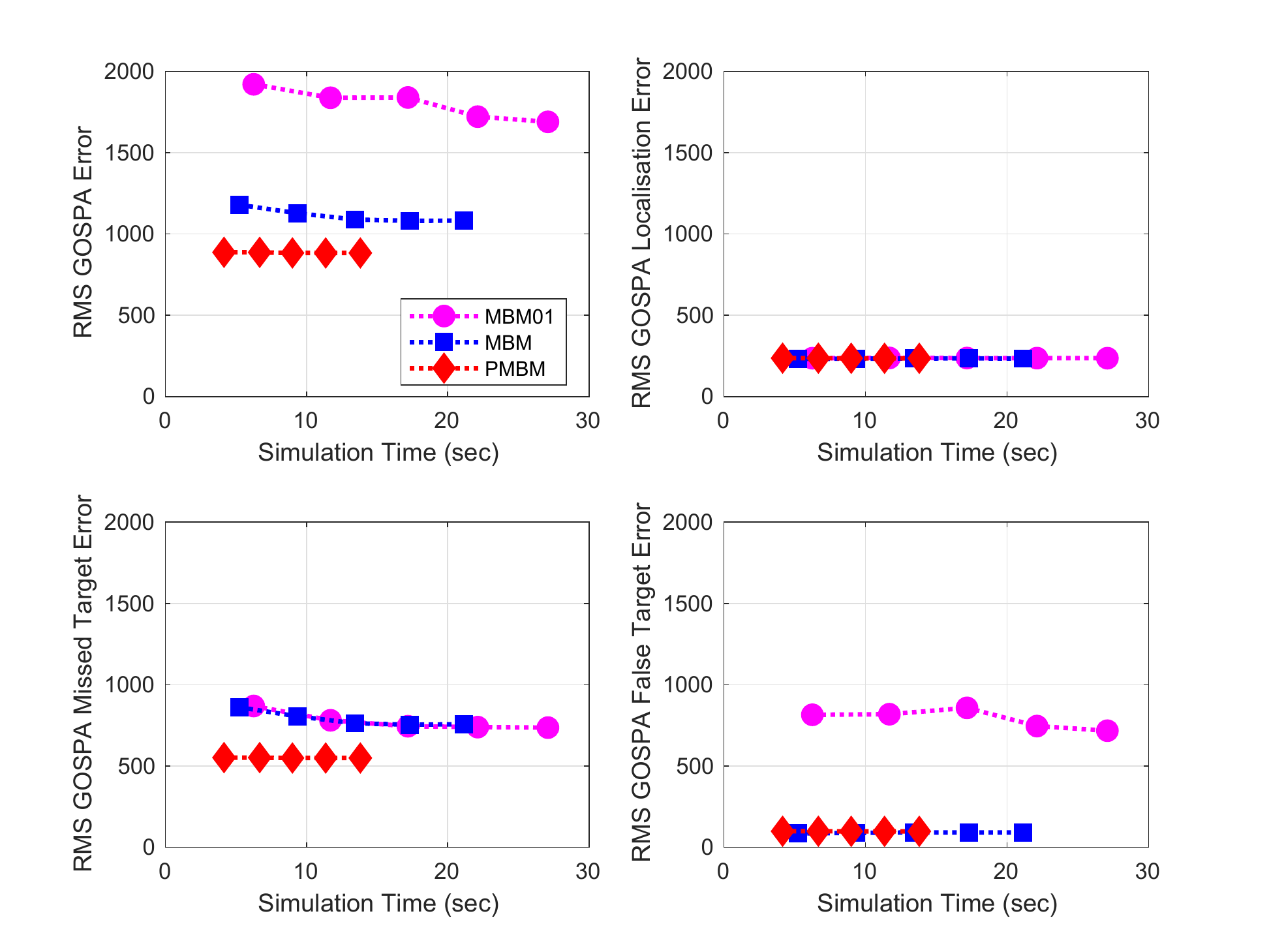}
\par\end{centering}
}
\par\end{centering}
\centering{}\subfloat[Scenario 3]{\begin{centering}
\includegraphics[width=1\linewidth]{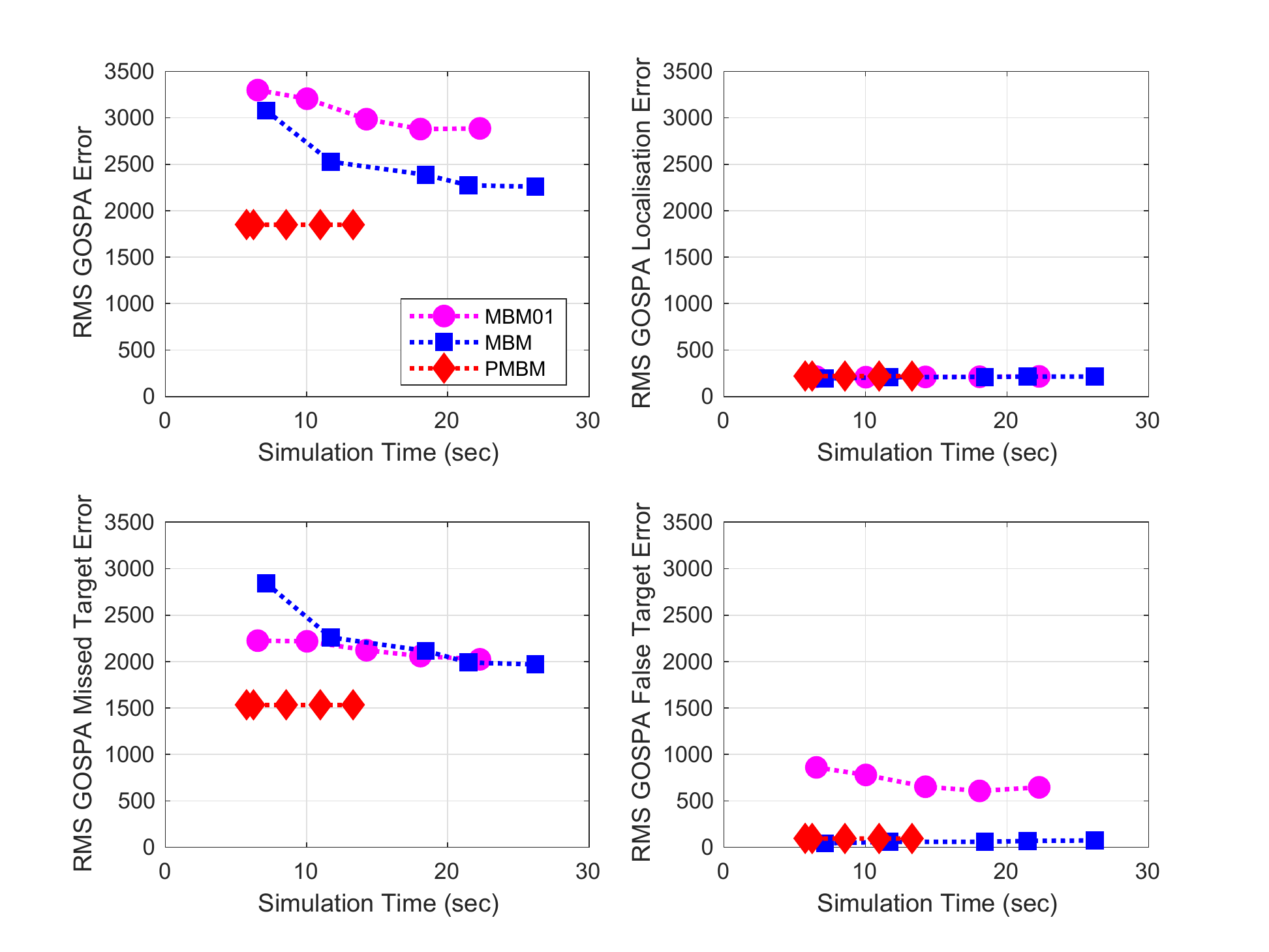} 
\par\end{centering}
}\caption{\label{fig:results}Performance comparison among the PMBM filter,
the MBM filter and the $\text{MBM}_{01}$ ($\delta$-GLMB) filter:
simulation time versus mean square GOSPA error and its decomposition.
For the same filter in each sub figure, the scatter points from left
to right marked with the same color correspond to $N_{h}$ equal to
$100,200,300,400$ and 500.}
\end{figure}

We perform 100 Monte Carlo runs and obtain the average root mean square
GOSPA error ($p=2$, $c=10$, $\alpha=2$) as well as the average
running time, summed over 81 time steps for each algorithm, as shown
in Figure \ref{fig:results}. For the PMBM filter and the MBM filter,
target states are extracted from Bernoulli components, contained in
the MB with the highest weight, whose existence probability is above
0.4. This estimator allows two consecutive misdetections for $p_{D}=0.9$
and $p_{S}=0.99$ to report an estimate. As for the $\text{MBM}_{01}$
filter, target states are extracted from the Bernoulli components
in the MB with maximum a posteriori cardinality and highest weight.

From the simulation results, we can see that the PMBM filter has the
best filtering performance in terms of GOSPA error and computational
time. The $\text{MBM}_{01}$ filter has significantly larger false
detection error than the PMBM filter and the MBM filter. We found
that this is because the filter $\text{MBM}_{01}$ usually fails to
report the death of the blue target around midpoint, where targets
are all in close proximity and data association becomes highly ambiguous.
This observation confirms the fact that MBM parameterisation can represent
the true posterior better than the $\text{MBM}_{01}$ parameterisation.

The MBM presents larger missed detection error than the PMBM filter
in Scenario 1, where we have an informative birth process. This difference
becomes larger when we have a broad birth prior density. This is
a drawback of having an MB birth density with identical Bernoulli
components, which might result in additional data association uncertainty
when associating measurements to Bernoulli birth components.

The advantage of having a Poisson point process birth over a multi-Bernoulli
birth can be clearly seen from the simulation result of the third
scenario, in which the PMBM filter has considerable smaller GOSPA
error than filters using multi-Bernoulli birth. When prior birth
information is vague, it is especially advantageous to use a measurement
driven approach in which new Bernoulli components are created from
measurements, as in the PMBM filter.

\section{Conclusions\label{sec:Conclusions}}

We have proposed a Gaussian implementation of the MBM filter using
Murty's algorithm to prune the global hypotheses. The MBM filter is
a special case of the PMBM filter \cite{Williams15b,Angel18_b} that
arises when the birth model is multi-Bernoulli or a mixture of multi-Bernoullis.
The MBM filter can be labelled if desired, and labelling does not
change the filtering recursion. Our simulation results indicate that,
among the two filters that use multi-Bernoulli birth model, MBM and
$\text{MBM}_{01}$, the MBM filter is superior. This is due to the
fact that the way of handling global hypotheses is more efficient.
However, PMBM outperforms both MBM and $\text{MBM}_{01}$ in the considered
scenarios. The Gaussian implementation has been provided for Gaussian/linear
models, but it can also be extended to nonlinear models using non-linear
Kalman filters \cite{Sarkka_book13}. 

The PMBM, MBM and $\text{MBM}_{01}$ filters can be extended to sets
of trajectories to provide full information on the trajectories followed
by the targets from first principles. The $\text{MBM}_{01}$ filter
(including its labelled version) and PMBM filter for sets of trajectories
were introduced in \cite{Granstrom18,Angel15_prov}, and the corresponding
MBM filter for sets of trajectories is also a special cases of the
PMBM, by considering a multi-Bernoulli birth process. Full details
of this filter will be provided in future work.

\bibliographystyle{IEEEtran}
\bibliography{12E__Trabajo_Angel_Mis_articulos_Finished_Fusion_2019_MBM_filter_Referencias_fusion19}

\end{document}